\documentclass[11pt]{article}
\usepackage{fullpage}
\usepackage{natbib}
\usepackage{epsfig}
\usepackage{amsmath}
\usepackage[footnotesize]{caption}
\usepackage{setspace} 
\usepackage{lineno}

\begin{document}

\title{Implications of Dispersal and Life History Strategies for
  the Persistence of Linyphiid Spider Populations}
\author{Dr. Leah R. Johnson\\
{\small Statistical Laboratory/CMS, University of Cambridge}\\
{\small Wilberforce Road, Cambridge, CB3 0WB, United Kingdom.}\\
{\small Current address:  Ecology, Evolution, and Marine Biology}\\
{\small University of California Santa Barbara, Santa Barbara, CA, 93106}  
\\{\small E-mail: {\tt lrjohnson@lifesci.ucsb.edu}}
\date{}}

\maketitle

\begin{abstract}
  Linyphiid spiders have evolved the ability to disperse long
  distances by a process known as ballooning. It has been hypothesized
  that ballooning may allow populations to persist in the highly
  disturbed agricultural areas that the spiders prefer. In this study,
  I develop a stochastic population model to explore how the
  propensity for this type of long distance dispersal influences long
  term population persistence in a heterogeneous landscape where
  catastrophic mortality events are common. Analysis of this model
  indicates that although some dispersal does indeed decrease the
  probability of extinction of the population, the frequency of
  dispersal is only important in certain extremes. Instead, both the
  mean population birth and death rates, and the landscape
  composition, are much more important in determining the probability
  of extinction than the dispersal process. Thus, in order to develop
  effective conservation strategies for these spiders, better
  understanding of life history processes should be prioritized over
  an understanding of dispersal strategies.
\end{abstract}

\section{Introduction}

Dispersal strategies occur over both short and long spacial scales. At
all scales, it has been suggested that dispersal is a bet-hedging or
risk-spreading strategy used by organisms to deal with heterogeneous,
stochastic environments
\citep{courtney:1986,hopper:1999,kisdi:2002}. However, dispersal and
movement by individuals also have more concrete consequences for
populations. It allows them to utilize new resources and areas, it
connects separate populations within a metapopulation, and it may help
maintain population and metapopulation stability and decrease
extinction risk \citep{hanski:disp.chpt20,hansson:1991}. Dispersal
into novel environments can also result in local adaptations and
speciation \citep{clobert}.

Linyphiid, or money, spiders are one example of an animal that employs
both short and long distance dispersal strategies
\citep{thomas:1990}. For money spiders, long distance dispersal occurs
as a mostly passive process known as ballooning
\citep{duffy:1998}. During ballooning, the spider is able to float
within air currents, suspended by a single strand of silk. Nearly all
Linyphiid species have been observed ballooning, although ballooning
propensity varies between species \citep{thomas:1990,duffy:1998}.
Although it is unknown how far a spider can travel by this method,
observations of spiders ballooning over the ocean, far from land
\citep{darwin:1906}, place some bounds on what is possible.

Linyphiid spiders prefer to live in agricultural areas, such as field
or pasture land, where they predominantly feed on aphids, although
some species are generalist predators \citep{sunderland:1986}. Since
they are able to balloon into areas that have been disturbed by
agricultural processes, it has also been suggested that Linyphiid
spiders may be important for controlling outbreaks of pests in these
areas \citep{sunderland:1986,thorbek:2005}. However, the spiders are
themselves sensitive to agricultural activities, such as harvesting or
pesticide applications \citep{thomas:1997}. Additionally, since the
early 1970s, observations indicate that populations of many Linyphiid
species have been decreasing, possibly due to climate change
\citep{thomas:preprint}. Since this decline seems to correlate most
strongly with a reduction in days where the weather is appropriate for
ballooning, the difference in population outcomes between species may
be related to differing dispersal propensities
\citep{thomas:preprint}. More specifically, particular dispersal
strategies may allow some species to better cope with the agricultural
landscape, which is characterized by a heterogeneous environment and
fairly frequent high-mortality ``catastrophes''. However, it is
difficult to observe the details of both dispersal and life
histories of the spiders directly, so another approach is needed.


Various models of spider ballooning have been
developed. \citet{humphrey:1987} first developed a simple one
dimensional fluid dynamics model of a single spider, and more recently
\citet{reynolds:2006} proposed a stochastic model of the process in a
turbulent flow. \citet{thomas:2003} proposed a statistical model for
the distances travelled by money spiders in different weather
conditions parameterized with data from observations of spiders
collected during ballooning. This model indicates that these spiders
may be able to travel more than 30 km within a single day
\citep{johnson:2008}, which is within observed bounds.  However, none
of these models address the population consequences for this kind of
very long distance dispersal.

There have been previous models that have been constructed to address
how dispersal strategies interact with life history strategies and
field disturbances to influence Linyphiid population levels.
\citet{thorbek:2005} developed a very detailed Individual Based Model
(IBM) for one Linyphiid species, Erigone atra, within a two
dimensional landscape. Their model includes details of landscape
dynamics (including crop growth and weather, as well as different
types of disturbance), stage structured life histories, and
environmentally cued dispersal. They primarily focus on how variation
in specific landscape activities (such as crop rotation) and landscape
compositions effect population sizes. However, the detail and
specificity of this model has drawbacks. Many of the conclusions may
not be generalizable to other species, and the shear complexity and
computational power needed for this type of model can make exploration
of the possible behaviors of this system much more
difficult. \citet{halley:1996} developed a simpler one dimensional,
deterministic model of a spider population composed of ``dispersers''
and ``non-dispersers'' in an agricultural landscape. However, like the
\citet{thorbek:2005} model, the specificity of this model,
particularly the use of very specific deterministic disturbances,
makes it difficult to draw general conclusions about how
metapopulation persistence is impacted by factors such as dispersal
and life history strategies.

In this paper I examine a simple stochastic model of a metapopulation
of ballooning spiders within a heterogeneous environment. The primary
goal of the study is to understand how dispersal strategies impact
long term population persistence in the face of high levels of habitat
disturbance and mortality. I approach the problem in the spirit of a
population viability analysis
\citep{boyce:1992,coulson:2001,reed:2002}, determining how populations
characterized by different life history parameters and dispersal
propensities may be more or less likely to go extinct within 110 years
when faced with varying levels of catastrophic events. This time
horizon is used as it would be a reasonable time frame for
conservation targets. I begin by introducing the model in Section
\ref{sec:model}, followed in Section \ref{sec:sims} by the simulation
methods used to explore the model. In Section \ref{sec:cart}, I
introduce classification and regression trees (CART), which are used
to analyze the simulation output.  Results for a baseline case and
three variations are presented in Section \ref{sec:results}. Section
\ref{sec:disc} concludes the paper with a short discussion.

\section{Model Description} \label{sec:model}

The model presented here is comprised of three portions: a population
model with demographic stochasticity within an agricultural field; a
data driven dispersal model; and a stochastic environment,
incorporating field level catastrophes.  The model is formulated as a
quasi Individual Based Model (IBM). Whereas a full IBM would follow
each particular individual over their whole lifetime, here I employ a
novel approach whereby individuals are only followed during the
dispersal process -- population dynamics and catastrophes are not
individual-based. This approach has computational benefits, especially
when large numbers of individuals are being modeled.

The landscape considered in this model is comprised of a one
dimensional ribbon consisting of patches or fields, each of size $s$,
with periodic boundary conditions. Each simulated landscape consists
of 50 virtual fields, numbered sequentially from 1 to 50. All fields
are the same fixed size, $s=1.3$km, and have the same fixed carrying
capacity $K=300$ (i.e., I assume that carrying capacity is constrained
by space availability within a field \citep{halley:1996}).

\begin{figure}[h!]
\begin{center}
\includegraphics[scale=0.4, trim=0 40 10 10, clip=true]{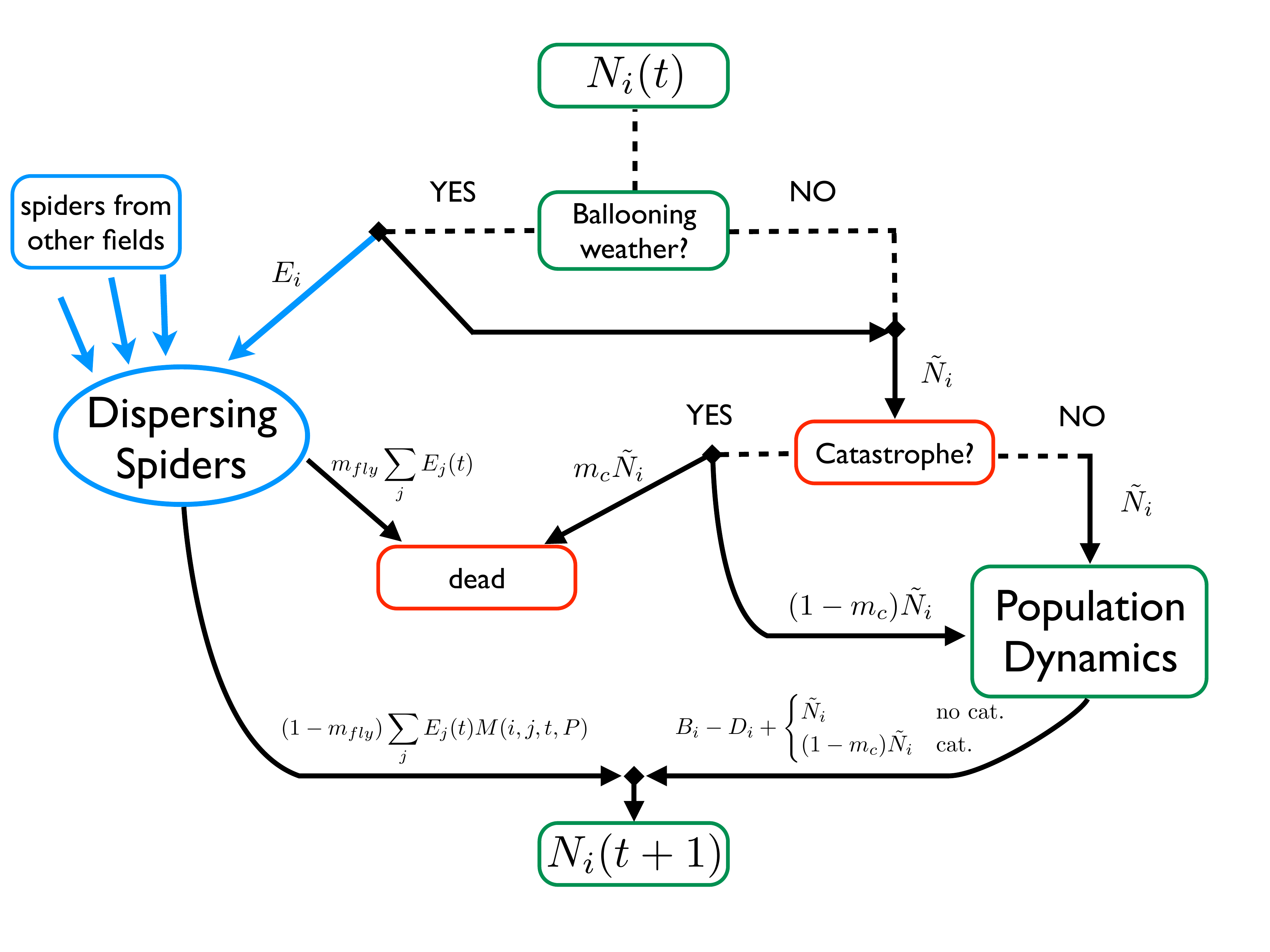}
\end{center}
\caption[]{Diagram of Model Flow}\label{f:model}
\end{figure}

In discrete time the number of individuals in the $i^{\mathrm{th}}$
patch changes as
\begin{equation}
N_i(t+1) = N_i(t) - D_i + B_i - E_i + I_i
\end{equation} 
where $D$ are the number of deaths in the patch, $B$ the number of
births, $E$ the number of emigrants leaving the patch, and $I$ the
number of spiders successfully immigrating to the patch. Figure
\ref{f:model} shows a diagram of the model flow at each time step,
which is explained in more detail below.

Each patch or field is one of $n$ types. Field types are 
characterized by their ``quality'', i.e., by the population birth and death rates
within the field. More specifically, spiders in high quality fields
could reproduce more quickly or are less likely to die from intrinsic
mortality than those in poor quality fields. Births and deaths are
modeled using simple stochastic logistic growth such that density
dependence acts to regulate reproduction and recruitment into the
adult population. In this case, at each time step a spider in field
$i$, of type $n$, dies with a probability $b_n$ and produces a single
(adult) offspring with probability $a_n\left(1-\frac{N_i}{\kappa_n}\right)$ where $\kappa$ is related to the traditional carrying capacity $K$ by $\kappa_n=a_nK_n/(a_n-b_n)$. In other words, I assume that only reproductive rates (and not death rates) are density dependent. Thus the expected number of births (here, recruited adults) and deaths in a
single field are given, respectively, by
\begin{align}
<B_i> & = a_n (\tilde{N}_i)\left(1-\frac{\tilde{N}_i}{\kappa_i}\right) \\
<D_i> &= b_n\tilde{N}_i 
\end{align}
where $\tilde{N}_i = N_i-E_i$ are the number of spiders that do not
disperse at time $t$.

Two hypotheses of dispersal are considered in the model: density
independent and density dependent. Thus the number of spiders
emigrating or dispersing from the $i^{\mathrm{th}}$ field at time $t+1$ can
either be a fixed proportion of the population in the patch at time
$t$,
\begin{equation}\label{eq:E1}
<E_i>\;=p_{\mathrm{disp}}N_i(t),
\end{equation}
or can vary with population density as
\begin{equation}\label{eq:E2}
<E_i>\;=p_{\mathrm{disp}}\frac{N_i^2(t)}{K_i},
\end{equation}
where $K_i$ is the carrying capacity in the $i^{\mathrm{th}}$ field,
and where we constrain $<E_i>\;\leq N_i(t)$. Thus, as $N_i(t)\rightarrow
K_i$ the proportion of individuals dispersing is the same for both the
density dependent and density independent cases. When $N_i<K_i$ spiders
are less likely to disperse when there is density dependent dispersal
than density independent dispersal, and vice versa for the case when
$N_i>K_i$. During dispersal, emigrants avoid mortality in the patch
but die with probability $m_{\mathrm{fly}}$. The mortality during
dispersal could include multiple factors such as predation or
desiccation. However, for simplicity here I assume a constant daily
mortality rate while dispersing. Spiders also cannot reproduce as they
disperse.

The number of spiders that immigrate into the $i^{\mathrm{th}}$ patch
is given by the sum of the spiders that leave all the fields ($E_j$),
survive dispersal, and consequently arrive in the $i^{\mathrm{th}}$
field. The dispersal kernel $M(i,j, t, P)$ describes the probability
that a spider starting in field $j$ lands in field $i$ on day $t$
given parameters, $P$. The data-driven model used to generate the
dispersal dynamics is presented in Section \ref{disp_mod}.

Spiders will only attempt to disperse under favorable weather
conditions. I assume that daily conditions are good for dispersal with
some fixed probability, $p_{\mathrm{fly}}$. In other words, out of
$\mathcal{D}$ days, the number of days with conditions favorable for
dispersal, $d$, is binomial with success probability
$p_{\mathrm{fly}}$: $d \sim \mathrm{Binomial}(\mathcal{D},
p_{\mathrm{fly}})$. Whenever conditions are favorable the numbers of
spiders that attempt to disperse are given by Equation (\ref{eq:E1})
or (\ref{eq:E2}), and when conditions are not favorable $E_i=0$ in
every field.

In the fields, ``catastrophes'', i.e., mortality events that wipe out
significant proportions of spiders in a particular field, can occur
\citep{thomas:1997}. A catastrophe with mortality rate $m_{c}$ occurs
on a given day with probability $p_c$. Catastrophes occur after
dispersal has begun (so that dispersing spiders can escape
catastrophes) but before births or (intrinsic) deaths. In addition,
all parameters that determine dispersal behaviors or population
dynamics are fixed and constant through time.

I am primarily interested in how variation of four parameters, given
the other parameters as fixed (see Table \ref{tb:params} and Section
\ref{disp_mod}), influences the probability of
extinction. Specifically I look at: the probability of catastrophe,
$p_c$; the probability of weather suitable for flying,
$p_{\mathrm{fly}}$; the probability that a spider disperses in good
weather, $p_{\mathrm{disp}}$; and the mortality rate experienced
during flying, $m_{\mathrm{fly}}$. In addition, the dispersal
probability can be either density dependent or density
independent. The catastrophe rate, $p_c$, is regarded here as being
primarily human induced mortality, for example due to application of
pesticides in a field. Two of the parameters, $p_{\mathrm{fly}}$ and
$m_{\mathrm{fly}}$, can be viewed as environmentally determined
parameters. The parameter $p_{\mathrm{fly}}$, which may be decreasing
for these spiders due to climate change \citep{thomas:preprint},
constrains the opportunities for dispersal into new habitats, and the
ability for spiders employing any dispersal strategy to escape local
mortality events. The mortality rate during dispersal,
$m_{\mathrm{fly}}$, includes mortality from various factors, such as
predation and desiccation. Thus the spiders must weigh the risks of
dispersing against the risks of catastrophes or benefits of
reproduction if remaining in a field. The final parameter,
$p_{\mathrm{disp}}$, together with the options for density dependence
or not, thus determine what I consider the evolved ``dispersal
strategy''.

\begin{table}
\begin{center}
\begin{tabular}{| c  c  c | }
\hline
Symbol & Description  & Value or Range\\
\hline
$k$ & number of fields/patches & 50 \\
$s$ & field size (km) & 1.3 \\
$n$ & number of types of fields & 5 \\
$K$ & field carrying capacity (spiders) & 300  \\
$m$ & dispersal time in each day (minutes) & $(0, 480)$ \\
$m_c$ & mortality level in a catastrophe & 95 \% \\
$b_i$ & per capita death rate in the $i^{\mathrm{th}}$ field & 0.02 \\ 
$a_i$ & per capita birth rate in the $i^{\mathrm{th}}$ field & [0.05, 0.2875, 0.525, 0.7625, 1] \\ 
\hline
$p_{\mathrm{fly}}$ & daily probability of suitable dispersal weather & $(0,1)$\\
$m_{\mathrm{fly}}$ & daily mortality rate in flight & $(0,1)$\\
$p_{\mathrm{disp}}$ & probability a spider disperses, given good weather & $(0,1)$\\ 
$p_c$ & daily probability of catastrophe, per field & $(0,0.5)$\\
\hline
($v_1$,$g$) & wind speed at 1m, wind speed
gradient & $(2.2, 1.25)$ \\ 
$E$ &  ascent/descent parameter & 2.5\\
$1/\lambda$ & mean waiting time (minutes) & 12 \\
\hline
\end{tabular} 
\caption[]{Parameters and their values or ranges in for the baseline
  simulations\label{tb:params}}
\end{center}
\end{table}

\subsection{The Dispersal Model} \label{disp_mod} 

\begin{figure}[h!]
\begin{center}
\includegraphics[scale=0.55, trim=0 42 0 10, clip=true]{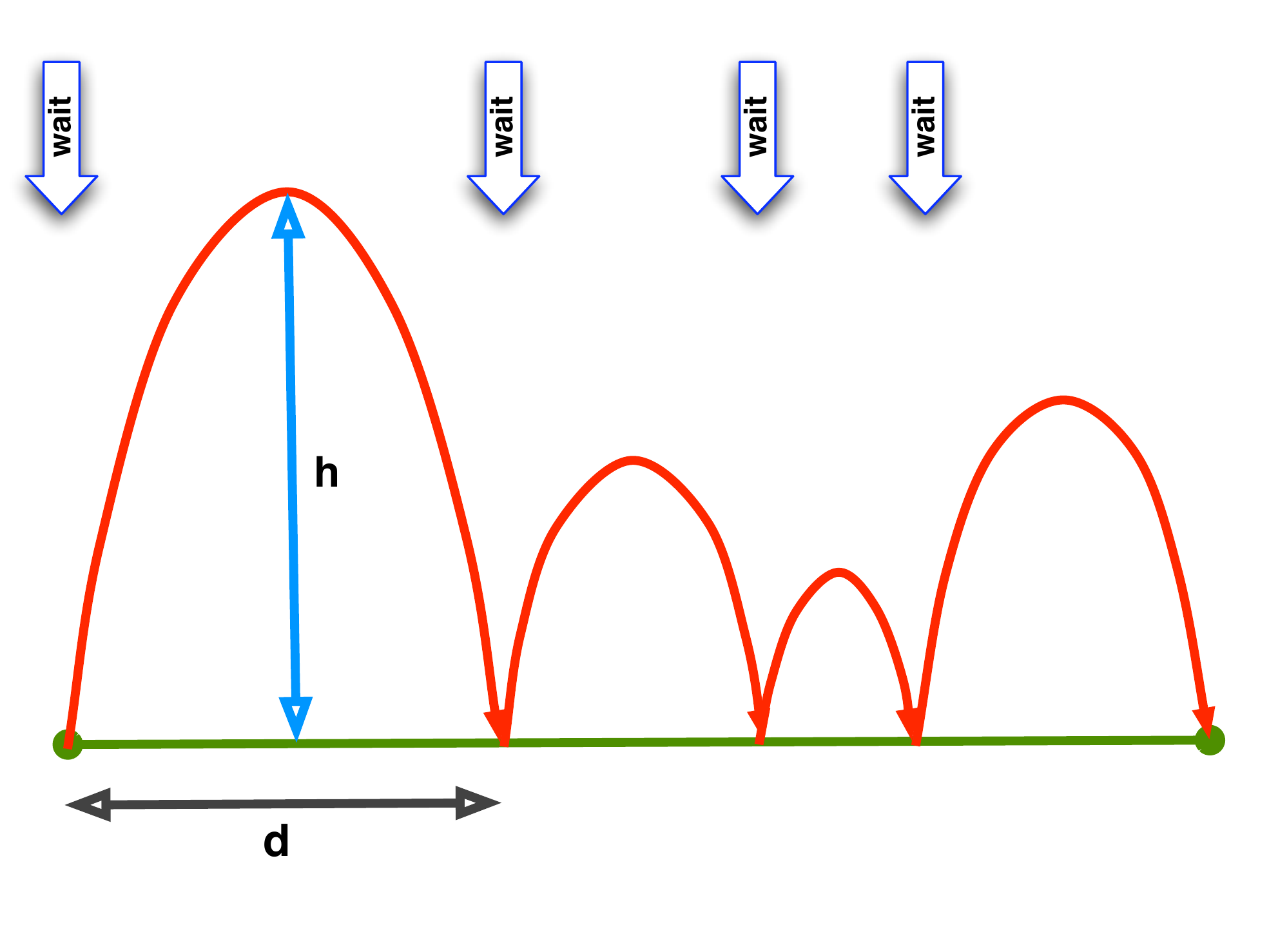}
\end{center}
\caption[]{Schematic representation of the dispersal
  process.}\label{f:disp}
\end{figure}

I utilize an established statistical model for the ballooning
dispersal kernel, parameterized with observational data
\citep{thomas:2003,johnson:2008}. For this approach, dispersal is
modeled as follows (see Figure \ref{f:disp}). Suppose that on a day
with appropriate weather for ballooning, there are $m$ minutes of good
weather for dispersal (specifically, take-off). At the beginning of
the day, each spider decides to begin dispersal attempts with
probability $p_{\mathrm{disp}}$. Upon attempting to disperse, the
spider first waits some time $\tau$ before successfully
ballooning. The waiting time has a exponential distribution so that
$\tau \sim \mathrm{Exp}(\lambda)$, where the mean waiting time (12
minutes, corresponding to $\lambda =\frac{1}{12}$) is fitted from
observational data. When the spider successfully takes off, it will
travel some distance $d$ during its flight. The distance is calculated
from the maximum height, $h$, that the spider can achieve in a flight,
with the distribution of heights given by
\begin{equation}
  f(h) = c_1 b(h^{-b} - h^{-(b+1)})
\end{equation}
for $1<h<h_{\mathrm{max}}$, where $h_{\mathrm{max}}$ is the maximum
possible height (here 1000m), $b$ describes changes of spider density
with height (from field data), and $c_1$ is a normalization
constant. Flights consist of an ascent at a fixed rate $\alpha$ up to
the maximum height, then a descent at rate $\delta$, the terminal
velocity. The horizontal wind speed varies with height and is
parameterized using field data. Single flight distances $l$ are
calculated by integrating wind speed for heights up to $h$,
\begin{equation}
l=
\begin{cases}
\frac{1}{E}\left(h v_1 +\frac{(h+1)\ln(h+1)-h}{g}  \right) & h\leq100\\
\frac{1}{E}\left(100 v_1 +\frac{(101)\ln(101)-100}{g}  \right) +
\left(\frac{h-100}{E} \right)\left(\frac{\ln(101)}{g}+v_1 \right) & 100<h<1000,
\end{cases}
\end{equation}
where $E$ is the ascent/descent parameter, defined as $1/E = 1/\alpha +
1/\delta$, and $v_1$ and $g$ are the wind speed at 1m and the wind speed
gradient, respectively. Both of these parameters are estimated from
observational data \citep{thomas:2003}.

\begin{figure}
\begin{center}
\includegraphics[scale=0.4, trim = 0 70 0 0, clip=true]{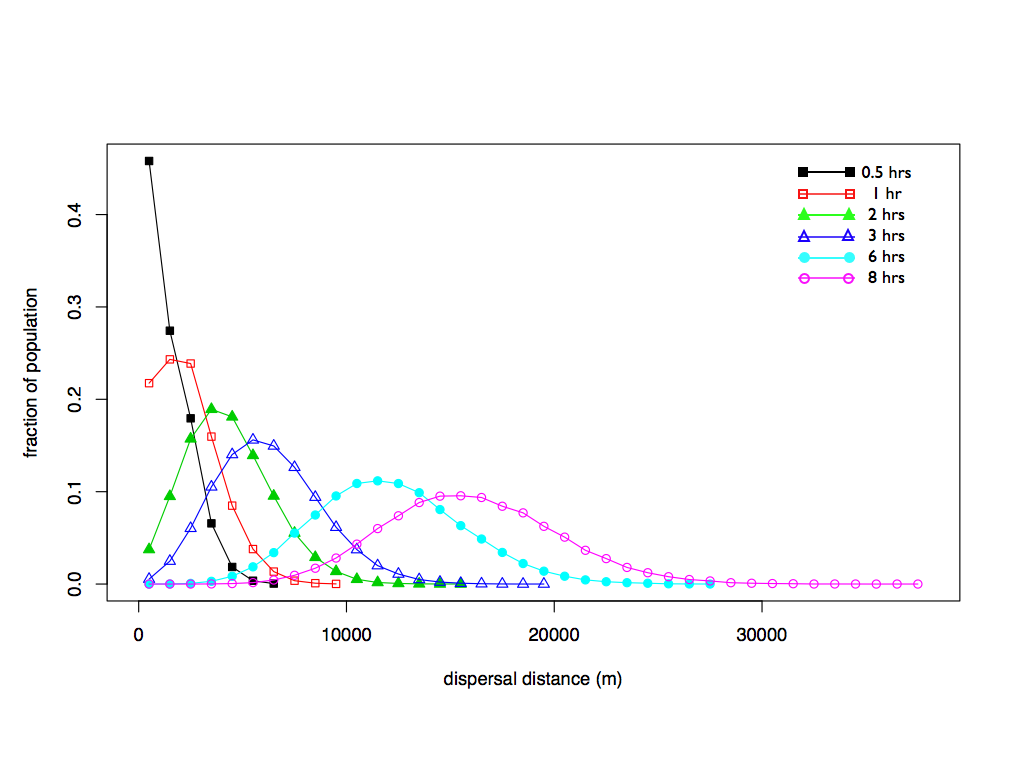}
\caption{Simulated distributions of total flight distances achieved by
  spiders during dispersal episodes of $m=\{0.5, 1, 2, 3, 6, 8\}$ hours
  with wind speed parameters $v_1 = 2.2$ and
  $g = 0.8$, and ascent/descent parameter $E = 2.5$. }
\label{f:disp2}
\end{center}
\end{figure}

During a day with $m$ minutes of good weather for dispersal, the
spider alternates waiting and flying (see Figure \ref{f:disp}). The
spider continues attempting to take-off throughout the entire time
available for flying, in order to travel the furthest distance
possible.  At the end of the day, it has travelled a total distance
$L$, and will generally be in a new field. Figure \ref{f:disp2} shows
simulated travel distances from this dispersal model for the flight
parameters used in the simulations. Note that in these conditions very
large dispersal distances are possible -- up to 30 km when there are 8
hours of appropriate weather. However, even with only half an hour of
good weather dispersal distances of 5-7 km are possible. This implies
that spiders are very likely to reach another field, since fields in
agricultural landscapes are generally less than 5 km in length
\citep{halley:1996}, and on especially fine days spiders dispersing
from a single field are likely to end up spread across a very wide
area.

We could approximate the dispersal kernel, $M(i,j,t,P)$, with
$P=\{h_a, h_b, E, m, \lambda\}$, described by this model via Monte
Carlo simulation of individual flights interspersed with waiting. In
many cases the number of simulations needed to approximate this kernel
is greater than the total numbers of individuals dispersing at any
given time in the simulation. Thus I instead directly simulate the
path of each dispersing spider individually to place them in a new
field at the end of the dispersal step.

\section{Model Parameterization and Simulations} \label{sec:sims}

For a model such as the one presented here the probability of
extinction within some fixed period of time cannot be found
analytically.  Instead a simulation approach must be used. More
specifically, simulations are performed to find the probability that
the meta-population, under specific dynamics and environmental
conditions described by a set of parameters, will go extinct within
110 years. 

Simulation parameters and their values are summarized in Table
\ref{tb:params}.  Birth and death rates are not well known for most
Linyphiid species. For the baseline case I assume a daily intrinsic
mortality rate of $b_i= 0.02$, which gives a mean lifespan of 50
days. This is lower than has been observed in controlled laboratory
settings, where spiders can live for more than 100 days, but seems
reasonable as first approximation to natural populations
\citetext{C.~F.~G.~Thomas, unpublished data}. In laboratory
experiments egg production levels and the proportion of eggs produced
which are viable are quite variable across food regimes and species
\citetext{C.~F.~G.~Thomas, pers.~comm.}. Thus the range of parameters
assumed for the per capita ``birth'' rates in the fields, $a_i$,
(here, rates or recruitment to adult class) are broad (see Table
\ref{tb:params}). At the extremes of this range, the expected number
of adult offspring produced by a single spider that lives for 50 days
in a single field ranges from 2.5 to 50. Thus the five field types
correspond to the five levels of $a_i$ specified in Table
\ref{tb:params}, since $b_i$ is fixed.

The set of parameters that determine the dispersal kernel are fitted
from observations of spiders on a single day, and are given in Table
\ref{tb:params}. Weather appropriate for ballooning tends to occur
mostly in the morning through to early afternoon
\citep{thomas:2003}. Thus I assume that the amount of time during the
day that is suitable for the initiation of dispersal, $m$, is
uniformly distributed from zero up to 8 hours.

Observed mortality levels for spiders during disturbances can vary
fairly widely. For instance, mortality levels of around 20\% from
residual toxicity from pesticide application \citep{thomasPHD}, or
56\% to $>95$\% during the application of insecticides or other
agricultural operations have been observed \citep{thomas:1997}. For
the simulations presented here, I assume that the morality level
during a catastrophe is constant between fields and over time, and is
set to the high end of the observed range (95\%) as a worst-case
scenario.


Each simulation experiment consists of both density dependent and
density independent dispersal cases. For each of these two cases, 500
parameter combinations of the four parameters of interest ($p_c$,
$p_{\mathrm{disp}}$, $p_{\mathrm{fly}}$, $m_{\mathrm{fly}}$) were
chosen as a Latin-hypercube sample (LHS)  \citep{mckay:LHC} to
efficiently explore the response within the parameter space. Thus, one
complete simulation experiment consists of 1000 parameter/density
combinations. In each simulation, the landscape composition (i.e., the
assigned ``type'' of each field in the ribbon) is randomly chosen from
an appropriate distribution. Initially, populations in all fields are
set to the carrying capacity. The simulation output consists of
binomial extinction/survival results, as well full population
trajectories for a random subset of the simulations.

Each simulation in the experiment is run for 400,000 virtual days
($\approx 110$ years) or extinction ($\sum_i N_i(t) \leq 1$),
whichever is first. For each parameter set, at least 2 runs are
performed with 10\% of the parameters randomly chosen for a third run,
for a total of approximately 2100 simulations. The experiment is
repeated for four cases, each of which I describe below.

\begin{figure}
\begin{center}
\includegraphics[scale=0.75, trim=10 0 10 10, clip=true]{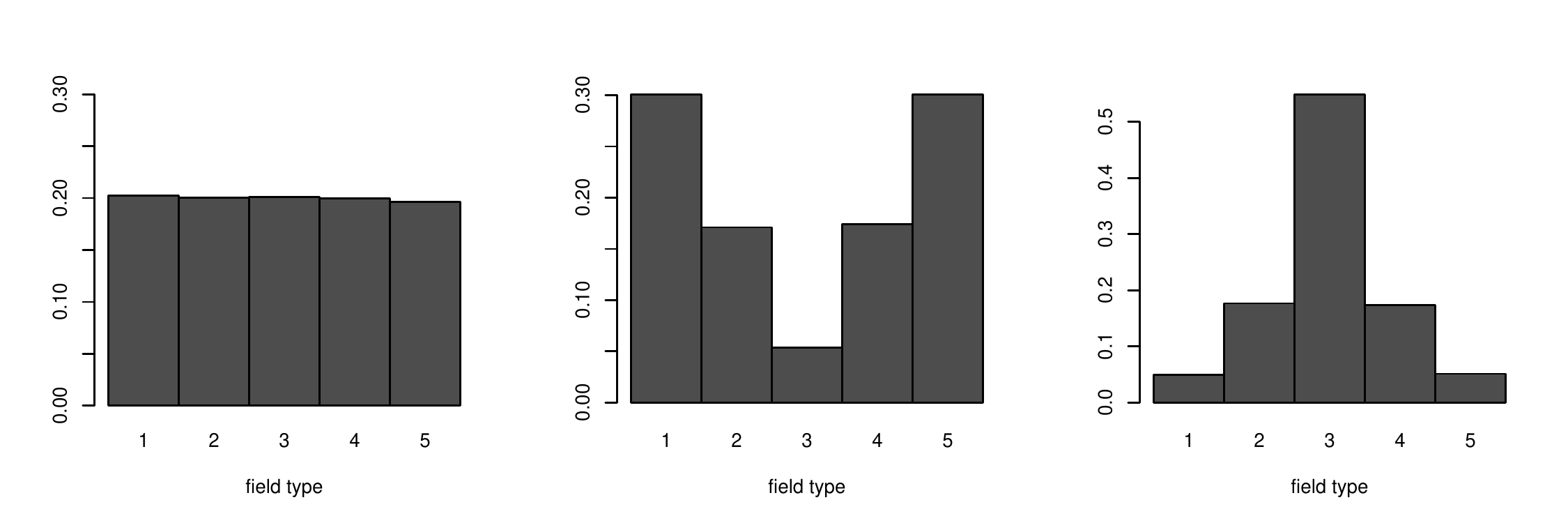}
\begin{tabular}{ccc}
\hspace{0.25in}  {\small (a)}\hspace{1.85in} & {\small (b)}
\hspace{1.85in}  & {\small (c)}
\end{tabular}
\end{center}
\caption[]{Draws from the multinomial distributions used to determine
  the field compositions in (a) the baseline main simulation discussed
  Section \ref{sec:bl} and (b) \& (c) the two variations discussed in
  Section \ref{sec:ls}}\label{fig:fields}
\end{figure}

\subsection{Baseline Simulations} \label{sec:bl}

The first of the simulation experiments is the ``baseline'' case. The
field parameters used for this set of simulations are given in Table
\ref{tb:params}. The landscape composition is determined by choosing
fields uniformly (Figure \ref{fig:fields}(a)) from the 5 possible
field ``types'', corresponding to the the 5 possible values of the per
capita birth rate, $a_i$, given in Table \ref{tb:params}.


\subsection{Baseline with High Variance in Field Composition
  (Best/Worst)} \label{sec:BW}

In the second simulation experiment, I begin to explore the effects of
changing the distribution of field types (relative proportions of high
vs.~low quality fields) in the landscape while leaving the types of
fields (the birth and death rates) the same as in the baseline
case. Whereas in the baseline case \ref{sec:bl} the distribution of
fields was uniform (Figure \ref{fig:fields}(a)), in the second
experiment I consider the case where most fields are either very good
or very bad, with fewer of intermediate quality, as shown in Figure
\ref{fig:fields}(b). This results in the same mean birthrate
throughout the fields, $\bar{a}=0.50$ as in the baseline
case. However, here the variance is higher, $\mathrm{Var}(a)\approx
0.156$. Throughout the rest of the paper I will refer to this case as
the ``Best/Worst'' scenario.

\subsection{Baseline with Low Variance in Field Composition (Many
  Moderate)} \label{sec:MM}

In the third simulation experiment, I again explore the effects of
changing the distribution of field types. However, in this case the
trend is the opposite of the Best/Worse scenario: there are many
moderately good fields, and fewer of both the low and high quality
fields (Figure \ref{fig:fields}(c)). Again this distribution has the
same mean birthrate throughout the fields, but has lower variance than
either of the previous two experiments, $\mathrm{Var}(a)\approx
0.0425$. In the discussion that follows, I refer to this experiment as
the ``Many Moderate'' scenario.

\subsection{Variation in Birth and Death Rates (Population
  II)} \label{sec:P2}

The three scenarios that I have described thus far all assumed the
same set of birth and death parameters in the fields, and only varied
the relative proportion of each. In the final experiment, I assume a
new set of birth and death rates. This could correspond to the same
species within a significantly different landscape, or a different
species within the previous landscape. Specifically, the within field
mortality level is increased compared to the baseline so that
$b=0.05$, and let the per capita birthrate in the $i^{\mathrm{th}}$
field is drawn from the set $a_i = [0.06, 0.1, 0.25, 0.3, 0.79]
$ 
That is, this population experiences two types of ``very low'' quality
fields, two ``low'' types, and one ``very good''. The distribution of
these fields within the landscape is uniform, i.e., the field type is
drawn from the multinomial distribution shown in Figure
\ref{fig:fields}(a). This set of birthrates has mean $\bar{a}=0.3$,
which is lower than the baseline case, and variance
$\mathrm{Var}(a)\approx 0.07$, which is intermediate compared to cases
already explored. In the analysis below I will refer to this
experiment as ``Population II''.

\section{Data analysis with CART} \label{sec:cart}

The simulation output (survival or extinction) can be treated as
binary data, and could be modeled in a number of ways. One traditional
statistical approach would be to look for the dependence of the
probability of extinction upon the parameters determining
environmental condition and dispersal strategy using a regression
approach, such as a generalized linear model (GLM)
\citep{mccullagh_nelder}. However, this approach is not appropriate
for these data, as the typical checks of the assumptions of additive
and homoskedastic error are not satisfied (data not shown). Instead, I
take a classification and regression tree (CART) approach
\citep{breiman}, which is a non-parametric statistical model that
allows for heteroskedastic errors and is applicable to both numerical
and categorical data.

In a GLM-type analysis we would appeal to the likelihood ratio test to
test for the significance of parameters, and proceed iteratively via
the forward/backward method to settle upon a final model. Building
trees proceeds in a similar, but usually non-iterative, manner. First
the Gini impurity \citep{breiman} maybe used to go forward and
``grow'' the tree (add splits/branches). Then cross-validation (CV) is
used to ``prune'' the tree back (remove splits/branches). All trees in
the following sections were fit using the {\tt rpart} package
\citep{rpart} in {\sf R} \citep{cranR}. In the {\tt rpart} package,
the degree of pruning is determined by a complexity parameter ($C_p$) \citep{mallows:1973} that may
be chosen by the one-standard-error rule \citep[][Section
7.10]{hastie:2001}, or other similar methods. For more details on
fitting treed models in {\sf R} or {\sf S} see \citet{MASS}.

One advantage of a CART-type model is that the interpretation of
(pruned) tree diagrams is fairly straightforward. The goal of the
diagram is to indicate for what values of various predictor variables
the model predicts a given probability of a response variable. In the
case explored here, the real-valued predictors are:
$m_{\mathrm{fly}}$; $p_c$; $p_{\mathrm{fly}}$; $p_{\mathrm{disp}}$;
and we have a binary class-type predictor \{dense, not dense\}. The
response variable is extinction in 110 years. Since the tree is built
interactively by finding the variable at each split that best explains
the variation in the response, earlier splits are generally the most
important for understanding the response. If a tree does not split on
a particular variable, and the variable is not correlated with the
variables that do appear, then knowing the value of the variable does
not significantly improve one's knowledge of the probability of
observing the response variable.

At each node in the tree, a Boolean expression is given, together with
some indication of a value or range of values for that parameter. For
instance in Figure \ref{fig:bigtree}, the first node is labeled as
$p_c<0.223$. The nodes dangling from the branches here are called the
children, or child nodes. The child nodes dangling from the left
branch of this top node operate on data satisfying this Boolean
expression; the child nodes dangling from the right branch violate
it. Thus in Figure \ref{fig:bigtree} the left half of the tree
corresponds to the cases where $p_c<0.223$ and the right half of the
tree corresponds to $p_c\geq 0.223$. Branches can spilt at the
children (for instance in Figure \ref{fig:bigtree}, the left child has
another spilt at $m_{\mathrm{fly}}<0.6712$), until the branches
terminate at ``leaves''. Each leaf in the tree diagram here is labeled
with the probability that the response variable (extinction) is true,
and the number of observations/simulations which lie in the portion of
the parameter space described by the branches leading to the
leaf. Thus, the values at the left-most leaf in Figure
\ref{fig:bigtree} indicate that the probability of the population
going extinct within 110 years, if $p_c<0.223$ and
$m_{\mathrm{fly}}<0.6712$, is $< 3$\%, based on 290 simulations.

\section{Results} \label{sec:results}

\begin{figure}
 \begin{center}
 \begin{tabular}{ccc}
   \includegraphics[scale=0.525, trim=20 30 0 0, clip=false]{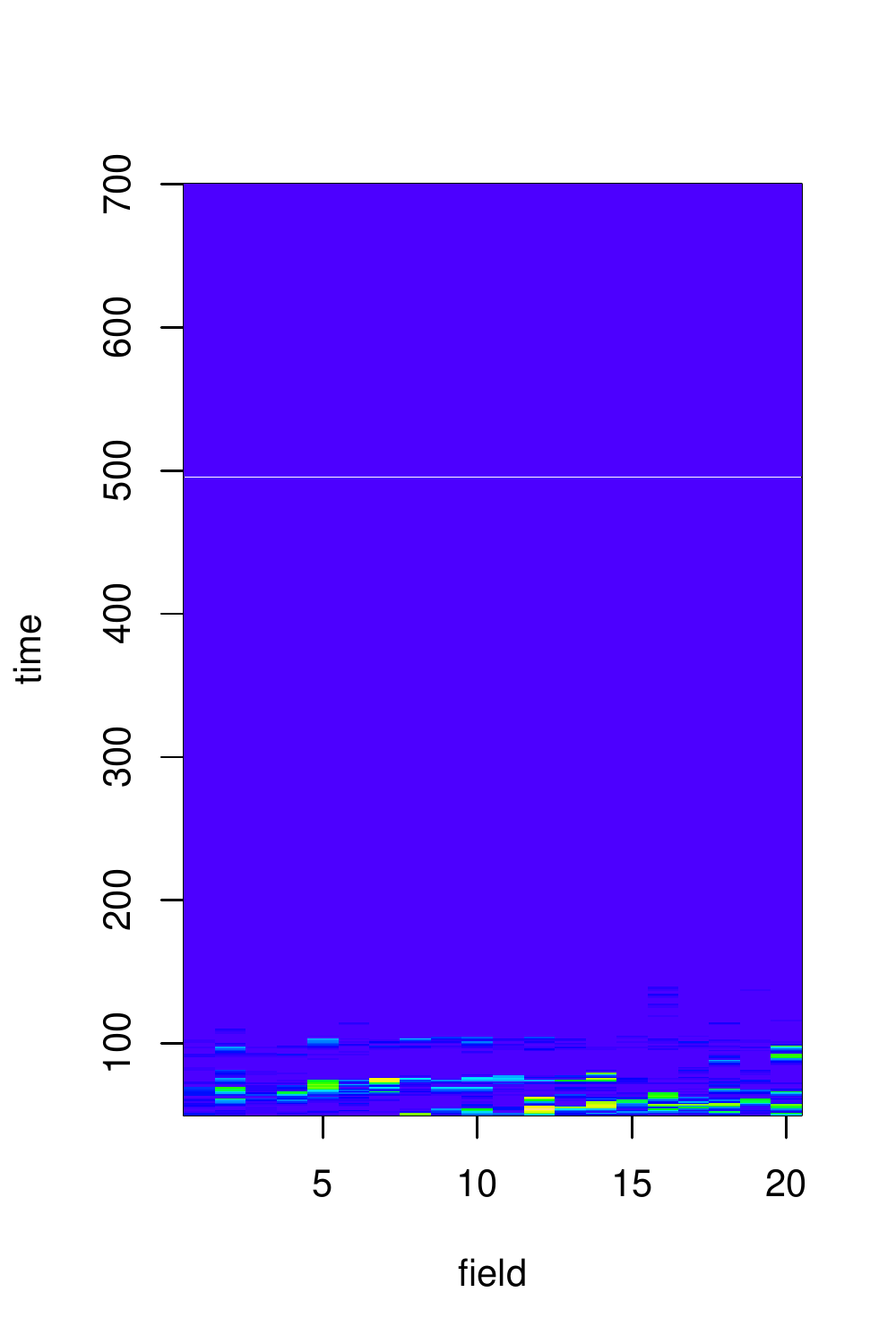} & 
   \includegraphics[scale=0.525, trim=20 30 0 0, clip=false]{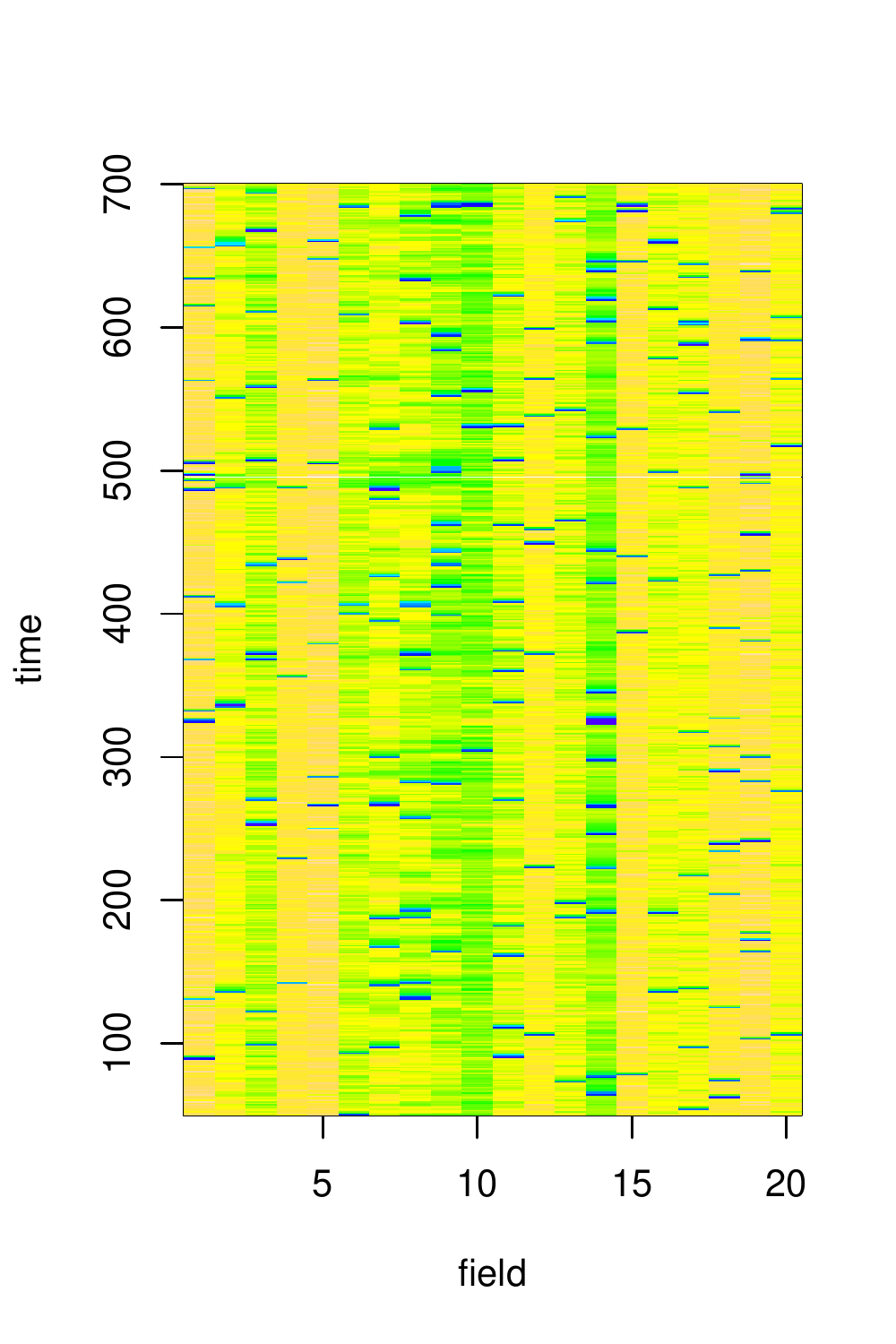}& 
   \includegraphics[scale=0.525, trim=20 30 0 0, clip=false]{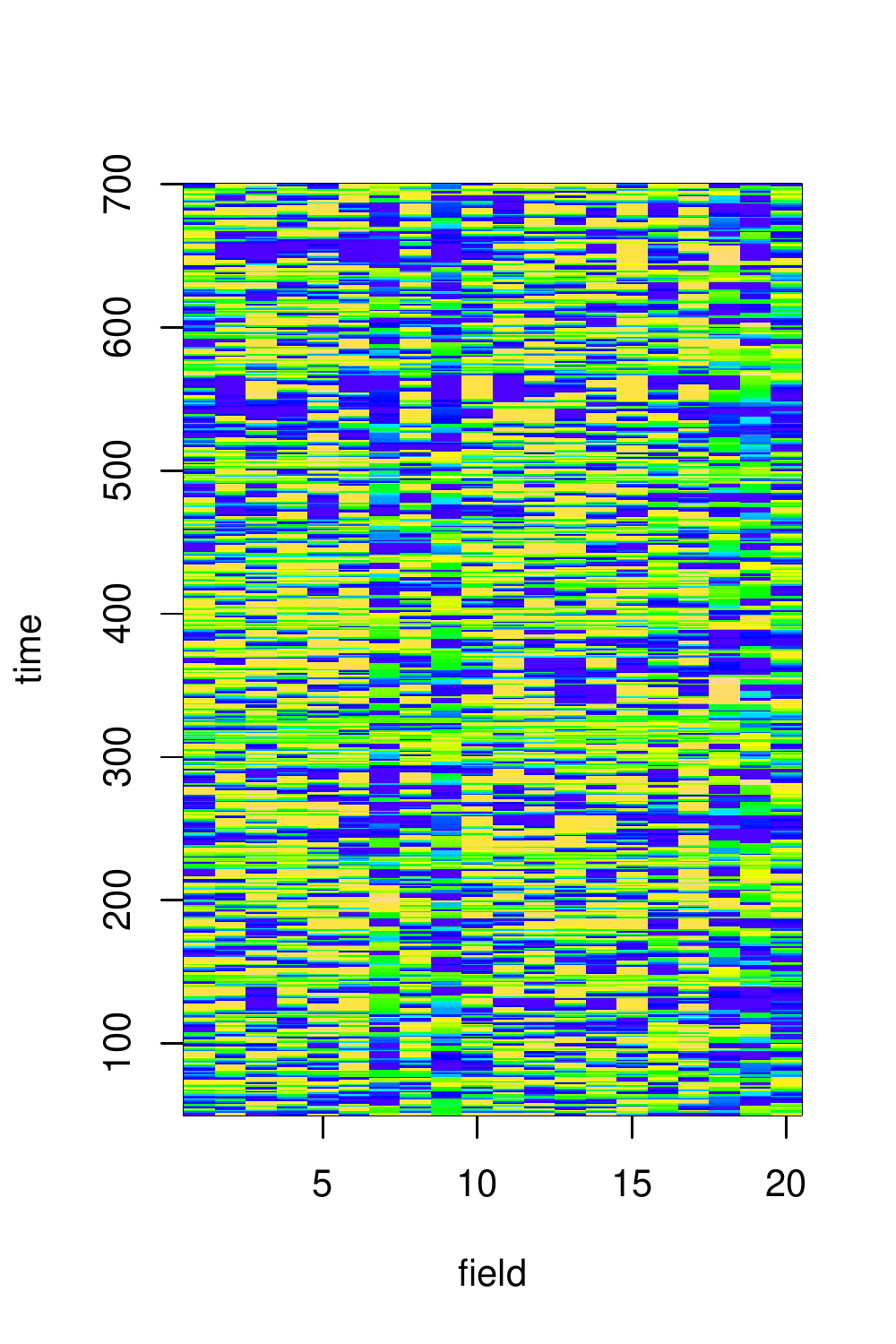}\\

 \includegraphics[scale=0.525, trim=20 0 0 0,
 clip=false]{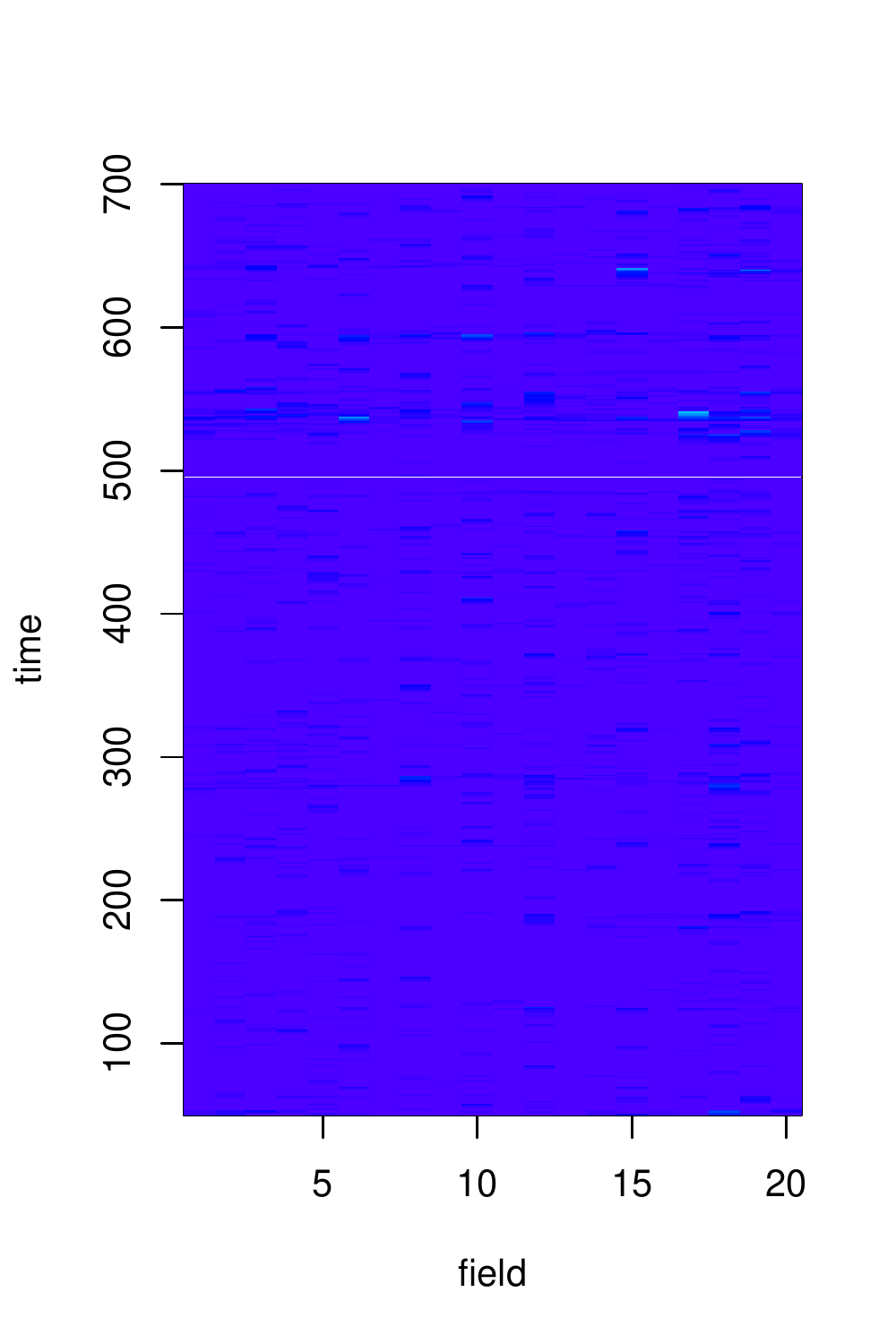}&
 \includegraphics[scale=0.525, trim=20 0 0 0,
 clip=false]{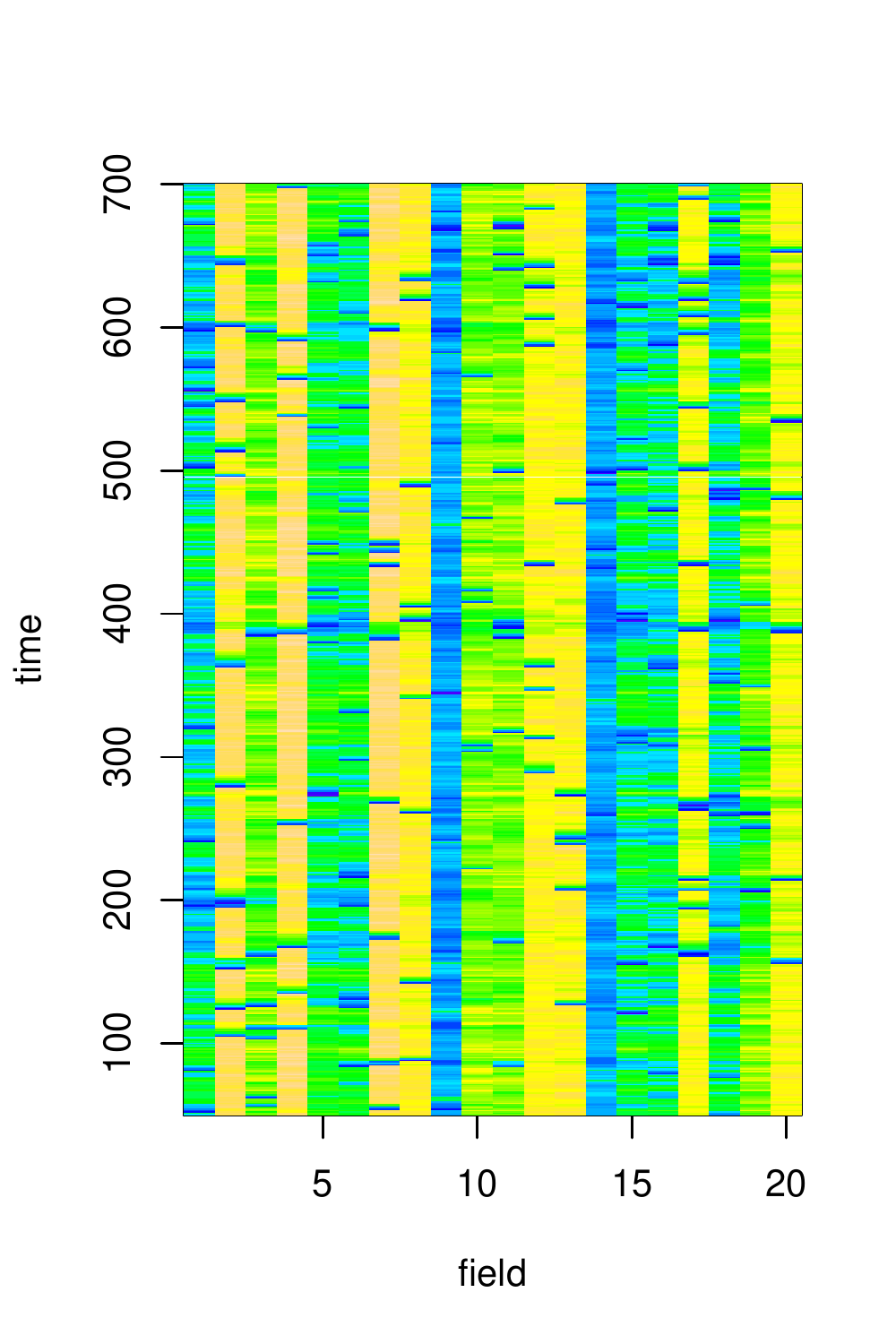}&
 \includegraphics[scale=0.525, trim=20 0 0 0,
 clip=false]{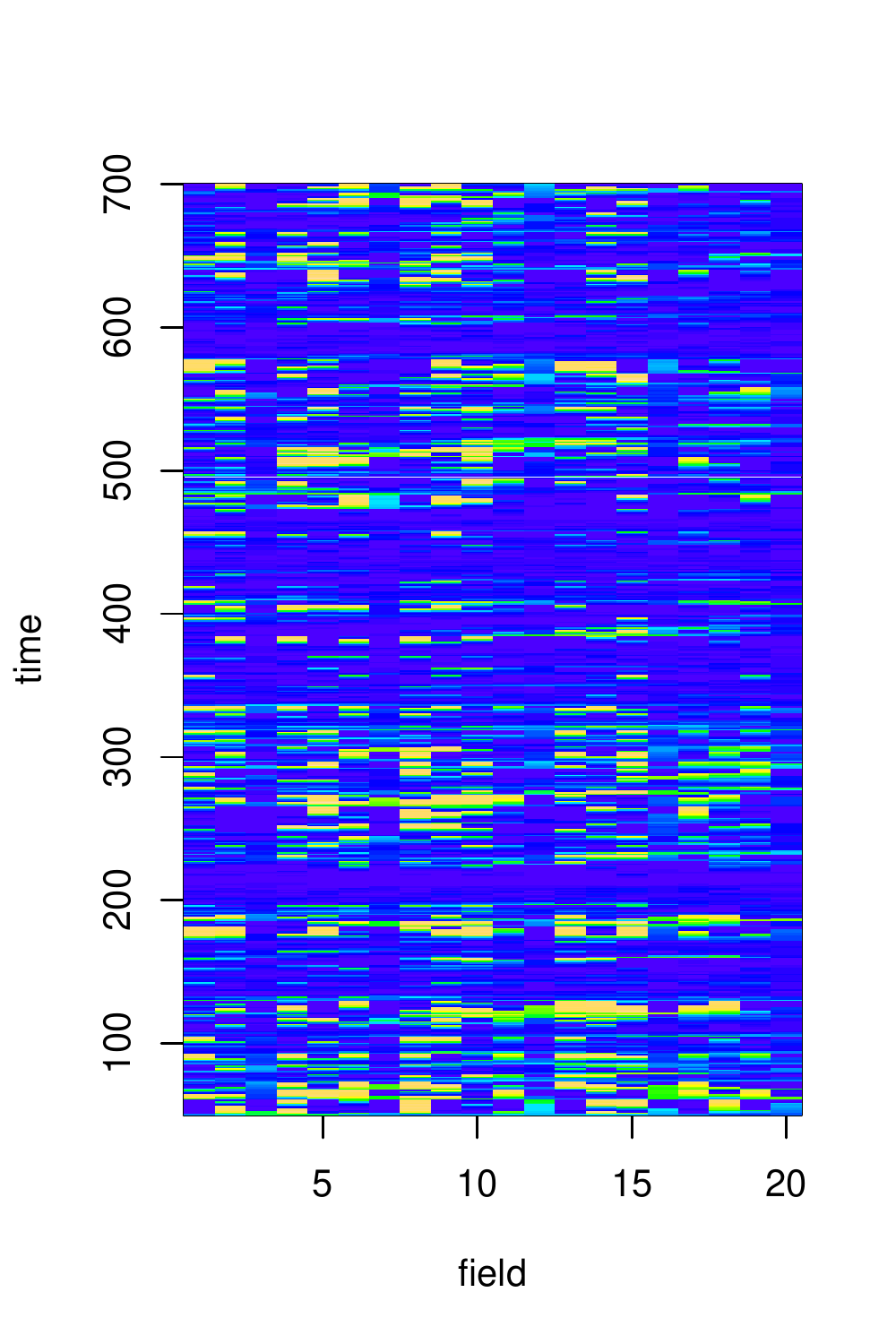}\\
 {\small (a)} & {\small (b)} & {\small (c)}
 \end{tabular}
 \end{center}
 \caption[]{Examples of density dependent (top) vs density independent
   (bottom) dispersal for three parameter settings with random field
   configurations. Fields are arranged in a line and are labeled by a location x-axis, and time on the
   y-axis. Violet indicates low or zero population size in a field
   and yellows larger populations. 
 } \label{fig:sims1}
 \end{figure}

Figure \ref{fig:sims1} gives examples of simulated population
 trajectories for a portion of the landscape (20 out of 50 fields)
 during the first 700 days. Results ranged from populations going
 extinct (or very nearly) early on (Figure \ref{fig:sims1}a) to very
 large populations across many of the fields (Figure
 \ref{fig:sims1}b). In simulations where populations were large and
 persistent, variation in field populations was due to differences in
 the habitat parameters in the different fields. Patterns tended to
 stabilize quickly, and most populations either went extinct very
 rapidly ($<100$ days, for example in Figure \ref{fig:sims1} (a), top), or
 persisted for the maximum duration explored here. In the remaining
 analysis I focus on the extinction/survival output.

\subsection{Baseline Scenario} \label{sec:bl.results} 

\begin{figure}
\begin{center}
\includegraphics[scale=0.4, trim=40 100 40 100, clip=true]{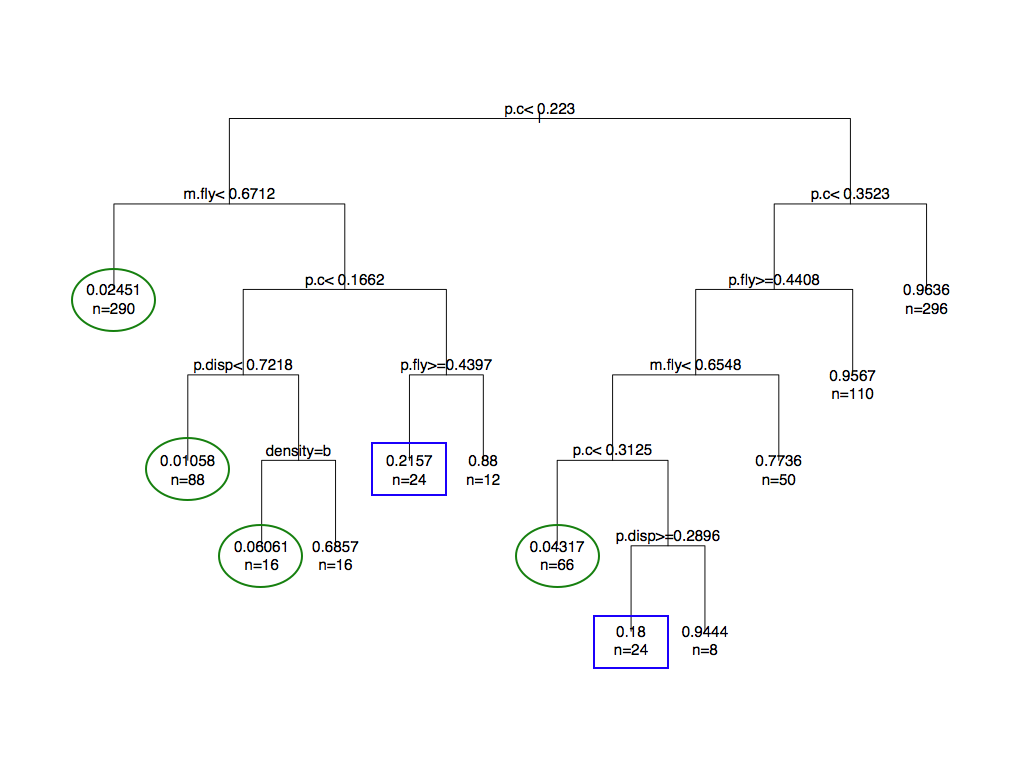}
\end{center}
\caption[]{Pruned tree showing the parameters and values that most
  determine the probability of metapopulation extinction within 100
  years for baseline scenario (Section \ref{sec:bl}). The splits from each node follow the
  rule left=TRUE. Density values of $\{a,b\}$ indicate
  density independent and dependent dispersal,
  respectively. Leaves with extinction probability of $<0.1$ are indicated with circles, and those with $<0.4$ with squares.}\label{fig:bigtree}
\end{figure}

I begin with the baseline case, as described in Section
\ref{sec:bl}. The pruned tree for the full data set is shown in Figure
\ref{fig:bigtree}. Even with the appropriate
pruning 
the tree is fairly complicated. First, we notice that the probability
of catastrophe, $p_c$, is the most important determinant of extinction
probability, as the initial branching depends on $p_c$, and there are
more branchings in the tree that depend on $p_c$ than any other
factor. The tree can be approximately viewed as having three regions
with low ($p_c<0.22$), medium ($0.22<p_c<0.31$), and high ($p_c>0.31$)
probability of catastrophe, that loosely correspond to low, medium,
and high probability of extinction within 110 years.

For most of the parameter space, the particular dispersal strategy
employed, (i.e., the probability of dispersing given good weather,
$p_{\mathrm{disp}}$, and density dependent or independent dispersal)
is not particularly important. When the risk of catastrophe is low, as
long as the inflight mortality is low enough
($m_{\mathrm{fly}}<0.67$), populations employing any dispersal
strategy are predicted to have a low probability of
extinction. However, when in-flight mortality is higher than this,
populations are only likely to persist when catastrophe levels are are
low ($p_c<0.1662$) and, simultaneously, either dispersal propensity is
not too high ($p_{\mathrm{disp}}<0.72$) or density dependent dispersal
is used (which reduces the effective dispersal propensity as long as
the population is below the carrying capacity). In other words, when
dispersal mortality is very high, frequent dispersal increases the
probability of population extinction, as one might expect.

At intermediate catastrophe levels (here $0.223<p_c<0.31$) populations
only persist in a very narrow range of circumstances where the
probability that a day has good weather for dispersal is greater than
44\% (more than 160 days per year) and, simultaneously, the inflight
mortality rate is not too high ($m_{\mathrm{fly}}<0.65$). In other
words, for the population to persist under intermediate disturbance,
there need to be adequate opportunities for at least some of the
spiders to disperse and survive to reach a new field. Outside of this
area of the parameter space, the probability of a population
persisting, particularly for catastrophe levels of over $0.3$, is very
low ($\approx$ 5-20\%) .


\subsection{Best/Worst and Many Moderate Scenarios} \label{sec:ls}

\begin{figure}
\begin{center}
\includegraphics[scale=0.4, trim=40 90 40 100, clip=true]{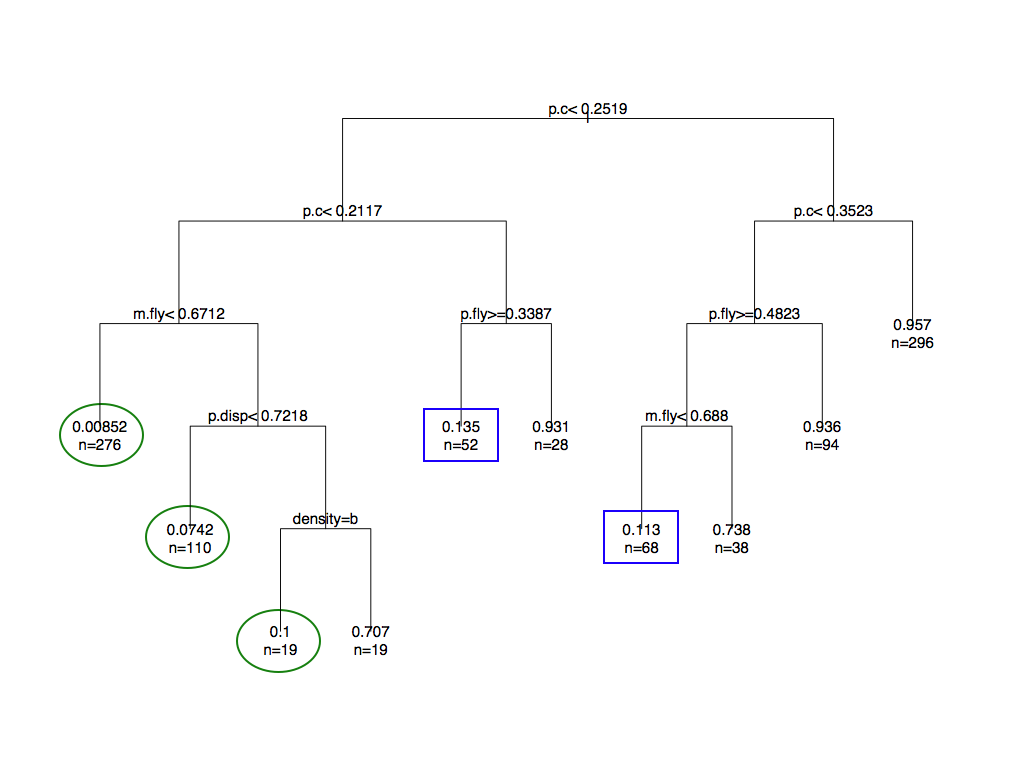}
\end{center}
\caption[]{Pruned tree showing the parameters and values that most
  determine the probability of metapopulation extinction within 100
  years for "Best/Worst" scenario.  (See Section
  \ref{sec:BW}) The splits from each node follow the rule
  left=TRUE. Density values of $\{a,b\}$ indicate density independent
  and dependent dispersal, respectively. Leaves with extinction probability of $<0.1$ are indicated with circles, and those with $<0.4$ with squares.}\label{fig:set2}
\end{figure}

Figure \ref{fig:set2} shows the results for the Best/Worst scenario,
which is characterized by high variance in the birthrates. First we
notice that, as in Section \ref{sec:bl.results}, the strategy
employed by the spiders is not particularly important for determining
the extinction probability (i.e., the tree only has a few splits, near
the leaves, that depend on either $p_{\mathrm{disp}}$ or
``dense''). Again the most important consideration is the level of
catastrophe. However, notice that in this case the population is
actually less sensitive to low levels of disturbance, where the
population has a very good chance of persisting (extinction
probabilities of $\approx 0.009$ to 0.1), as long as $p_c<0.21$. This
is likely due to the increased abundance of high quality fields. Like
the baseline case, the exception to this is when both in-flight
mortality and dispersal propensity are high ($m_{\mathrm{fly}}>0.67$
and $p_{\mathrm{disp}}>0.72$, respectively) and individuals utilize
density independent dispersal. In this case, populations have a
greater than 70\% chance of going extinct.


\begin{figure}
\begin{center}
\includegraphics[scale=0.4, trim=40 90 40 100, clip=true]{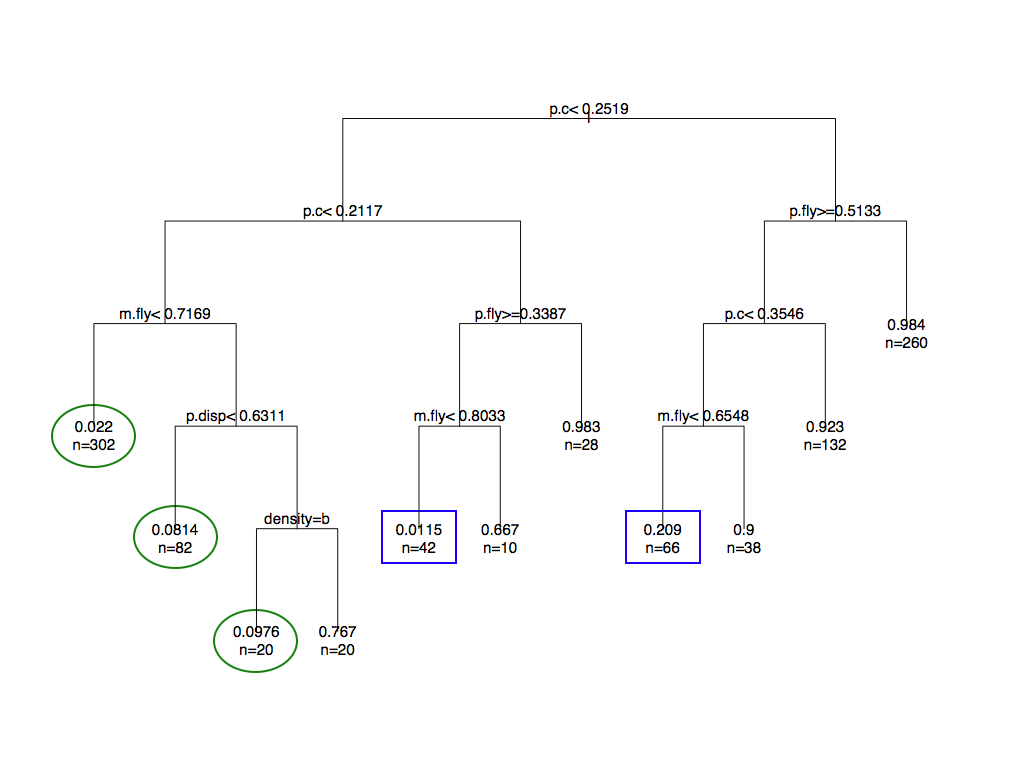}
\end{center}
\caption[]{Pruned tree showing the parameters and values that most
  determine the probability of metapopulation extinction within 100
  years for the "Many Moderate" scenario.  (See Section
  \ref{sec:MM}) The splits from each node follow the rule
  left=TRUE. Density values of $\{a,b\}$ indicate density independent
  and dependent dispersal, respectively. Leaves with extinction probability of $<0.1$ are indicated with circles, and those with $<0.4$ with squares.}\label{fig:set3}
\end{figure}

Figure \ref{fig:set3} shows the results for the Many Moderate
scenario, characterized by lower variance in the birthrate. The
resulting tree is nearly identical to the Best/Worst
scenario. However, in this scenario, populations are slightly more
likely to go extinct (extinction probability of $\approx 0.02$) when
both the probability of catastrophes and inflight mortality rates are
low ($p_c<0.211$, $m_{\mathrm{fly}}<0.7$) compared to the Best/Worst
case; instead the extinction probabilities more similar to the
Baseline. 


\subsection{Population II scenario} \label{sec:ls2}

\begin{figure}
\begin{center}
\includegraphics[scale=0.4, trim=40 90 40 100, clip=true]{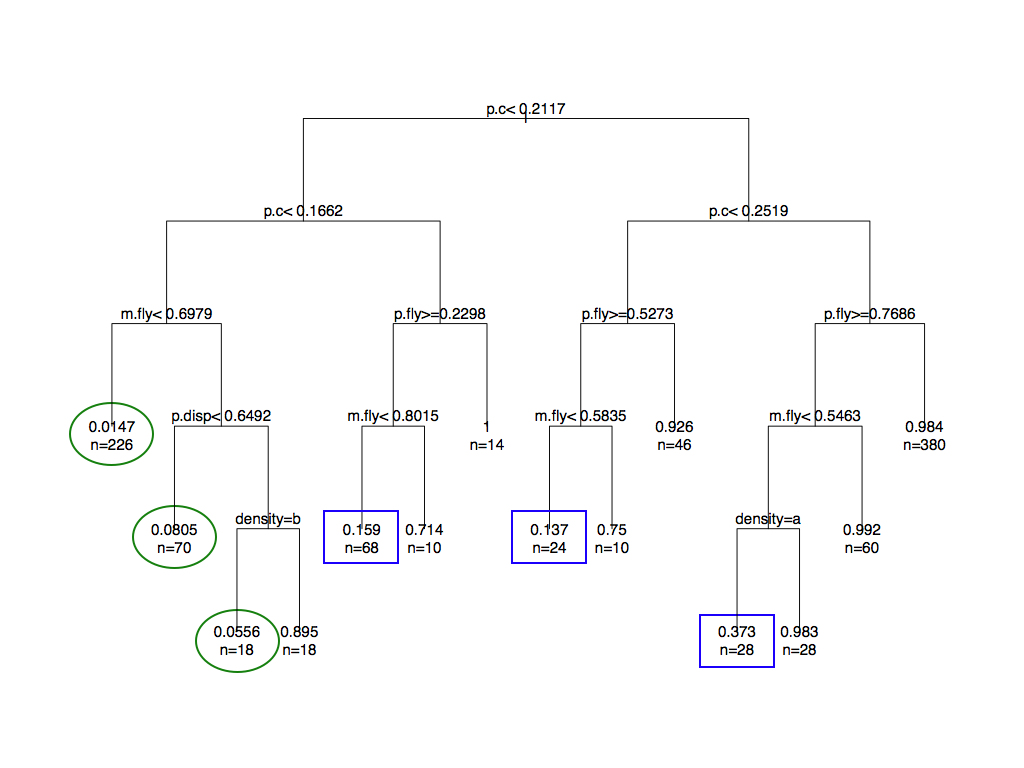}
\end{center}
\caption[]{Pruned tree showing the parameters and values that most
  determine the probability of metapopulation extinction within 100
  years for the Population II scenario, characterized by birthrate parameters $a=\{0.79, 0.3, 0.25, 0.1, 0.06\}$
  (mean birthrate of $\bar{a} = 0.3$, or 57\% of the baseline mean)
  and death rate parameter $b=0.05$ (death rate 2.5 times higher than
  the baseline) and with a uniform distribution of field types as
  shown in Figure \ref{fig:fields} (a). This .  (See Section
  \ref{sec:P2}) The splits from each node follow the rule
  left=TRUE. Density values of $\{a,b\}$ indicate density independent
  and dependent dispersal, respectively.  Leaves with extinction probability of $<0.1$ are indicated with circles, and those with $<0.4$ with squares.}\label{fig:set1}
\end{figure}

Figure \ref{fig:set1} shows the results for the final scenario
explored in this paper, the population II case. As one would expect,
since the intrinsic death rate is higher than the previous cases and
the mean birth rate is lower, much more of the parameter space results
in the extinction of the population. The threshold level of
catastrophe that results in a low probability of extinction is $p_c
\approx 0.17$, which is quite a bit lower than any of the previous
cases. Even with this low level of catastrophe, the population is
still only likely to persist if the dispersal mortality is low enough
($m_{\mathrm{fly}}<0.7$), the probability of dispersing on a day with
good weather is not too high ($p_{\mathrm{disp}}<0.65$), or density
dependent dispersal is utilized.

\subsection{Results Summary}

The results from all of the scenarios explored above show many
similarities. For instance, much of the tree structures, such as
primary splits depending on $p_c$, correlations between
$m_{\mathrm{fly}}$ and $p_{\mathrm{disp}}<0.65$, and density
dependence only being important in limited circumstances. In each
case, the probability of catastrophe determines the probability of
extinction within 110 years more than any other factor. The
similarities between the left-most branch in all four of the scenarios
also indicates that the product of in-flight mortality rate and
dispersal propensity, which together determine the expected proportion
of individuals within a field that will die on a day with good
conditions for dispersal, may be an important threshold for
determining extinction probability for a given catastrophe
level. However the quantitative results (especially locations of
splits) exhibit more variation. In particular, the results indicate
that the values of population parameters (birth and death rates) are
considerably more important for determining population persistence at
a given catastrophe level than the relative abundance of the different
types of fields, which in turn has a greater impact than changes in
the dispersal strategy (dispersal propensity and density dependence).

\section{Discussion} \label{sec:disc}


The results of the model presented in this study suggest that,
although a general dispersal ability is important for the persistence
and growth of Linyphiid spider populations, the exact details of this
dispersal strategy, i.e., whether dispersal is density dependent and
the particular probability of dispersing on a day with appropriate
weather, are less important than other factors in determining
persistence in the face of field level catastrophes. Instead, actual
catastrophe probability seems to be the most important factor in
determining the extinction probability, given landscape and life
history parameters. As demonstrated, one may observe thresholds in the
catastrophe level where the population switches from being very
unlikely to being very likely to go extinct. For instance, results for
populations with life histories and landscape distributions described
by the parameters in the baseline simulation suggest that if the daily
probability of a catastrophe is greater than 22\%, then there is
greater than 80\% chance of extinction within 100 years. Otherwise,
there is less than 10\% chance that the population would go
extinct. The baseline results also make it apparent that reducing the
catastrophe level further can help to mitigate the effects of
mortality during dispersal.

Although the model presented here fairly simple, it is able to capture
patterns that have been observed in more complex models. For instance,
the model developed by \citet{halley:1996} exhibited similar
thresholding behavior in population size/persistence with catastrophic
events, specifically landscape wide pesticide application (all fields
affected). They found that if all the fields were of the same type,
the population could persist (i.e., the population was $>0$) if the
field was sprayed no more than once per year with a pesticide that
caused 90\% mortality. By including a second field type that is less
ideal for habitat, but is not sprayed, the population remains large
even with higher frequency of pesticide application in other
fields. In the current study we find a similar increase in persistence
by limiting the average catastrophe rate across fields, instead of
explicitly including refuge habitats. This indicates that for highly
dispersive species, undisturbed land for refuges may not be as
necessary for population persistence as lower mean disturbance rates,
although providing refuges may be an efficient method for reducing the
mean disturbance rate. This is a similar result to one reported by
\citet{thorbek:2005} who found that some habitat needed to be
available for spiders at all times, although the habitat did not need
to be permanent.

On the other hand, by using a more simple model for some aspects, such
as the life history, I have been able to focus more on the more
general question of the relative importance of dispersal strategy
compared to other population and landscape factors for population
persistence in the face of catastrophes.  Although many of the
qualitative results of this model did not depend upon the life history
and landscape parameters, the quantitative predictions and, more
importantly, the threshold catastrophe levels do depend upon the
assumptions about the distribution of field quality in the landscape,
reproductive rates, and baseline mortality.  On the other hand, the
particulars of the dispersal strategy adopted by the spider (such as
density dependence or dispersal propensity) were not particularly
important under most circumstances. This is in contrast to
\citet{halley:1996}, who found a fairly strong dependence between
population size/extinction and the proportion of individuals
dispersing. This difference is could be due to a number of different
factors. One possibility is that this the effect of dispersal is less
apparent in the current model due to significant stochasticity in all
of the model processes. Another is that the difference could be an
effect of the stage structured population dynamics, which may result
in the particular amount of dispersal being more important in recovery
from a catastrophe. A third possibility is related to the fact that
the optimal proportion of dispersers in the \citet{halley:1996} study
was also strongly related to the proportion of non-habitat patches
within the landscape. This factor changes the risk of mortality while
dispersing, while simultaneously altering the population reproduction
parameters, and seems to be more important in determining the maximum
population level than the other factors they explored. A final
possibility is that the difference could also be due to the fact that
\citet{halley:1996} assume that each individual spider is either a
``disperser'' or ``non-disperser'' for its entire life-cycle. This
factor may also be part of why \citet{halley:1996}, and
\citet{thorbek:2005} draw conflicting conclusions about the effect of
field rotations on the population. If a portion of the population are
``non-dispersers'', then rotating a field would effectively increase
the catastrophe level for a large portion of the population, since
these individuals cannot escape a dramatic change in mortality due to
the rotation by dispersing. Although I do not deal with rotation
explicitly in this model, I expect it would have a similar, though
mild, effect here, as long as the rotations do not change the overall
distribution of fields in the landscape dramatically.


Since there is such a strong interaction between the effect of
population parameters and catastrophe level, the current study
suggests that the current patterns of decline are likely to be due to
a combination of both changing life histories and agricultural
practices (field composition and catastrophe level). In order to
preserve or increase spider populations in the future, we may want to
suggest conservation measures that seek to curb the levels of human
induced catastrophes in the environment. The observed thresholding
behavior in the model indicates that the development of a simple
guideline may be possible. For the parameters explored here the
thresholds were in the 20\% range. In other words, $\sim$ 20\% of
fields experience catastrophic mortality on a given day, and in a
single field we expect nearly 10 weeks worth of high mortality days
each year. Although pesticides applied to fields can remain toxic to
spiders for more than two weeks after application \citep{halley:1996},
and other types of disturbances also cause significant mortality
\citep{thomas:1997}, the predicted threshold seems to be fairly high.
However, this value depends fairly strongly on model parameters,
especially the population birth and death rates. Thus, more
observational data on the reproductive capabilities of target species
within various types of agricultural fields, and how these may be
affected by climate change, would be most useful for estimating this
threshold. Data gathered to estimate different dispersal
behaviors/propensities or changes in the proportion of days that are
suitable for ballooning would be less useful.

In the current simulations, the effect of a reduction in the number of
``habitat" fields has not been explored. This is partly because the
effect of reducing carrying capacity, $K$, on metapopulations is
fairly well understood \citep{mangel:1993a,mangel:1993b}. The addition
of ``non-habitat'' fields at random into the landscape, without
reducing the total $K$, would be equivalent to raising the level of
mortality during dispersal.

The current model focuses on the case where there are no
spatiotemporal correlations in either catastrophes or reproductive
schedules. It may be that these kinds of correlations could reduce the
tolerance of a population to disturbance, or make other dispersal
strategies, such as ones signalled by external factors, more
important. Including these factors this would be an important aspect
of future work.

\section{Acknowledgements}
L.~R.~J. was funded by BBSRC grant D20476 
as part of the National Centre for Statistical Ecology. Thanks to:
George Thomas for unpublished data and biological expertise on
Linyphiid life histories; Bobby Gramacy for advice on statistical
methods; Ian Carroll for comments on an earlier draft; and two very
helpful and thorough reviewers.



\footnotesize
\bibliographystyle{plainnat}


\end{document}